%% file: main.tex
\documentclass[sigconf]{acmart}

\acmConference[MSR 2024]{21st International Conference on Mining Software Repositories}{April 2024}{Lisbon, Portugal}

\AtBeginDocument{%
  \providecommand\BibTeX{{%
    \normalfont B\kern-0.5em{\scshape i\kern-0.25em b}\kern-0.8em\TeX}}}
    
\def\BibTeX{{\rm B\kern-.05em{\sc i\kern-.025em b}\kern-.08em
    T\kern-.1667em\lower.7ex\hbox{E}\kern-.125emX}}

\usepackage{amsmath,amsfonts}
\usepackage{array}
\usepackage{booktabs}
\usepackage[skip=1pt,labelfont=bf]{caption}
 \usepackage{calligra}
\usepackage{color, colortbl}
\usepackage{courier}
\usepackage{csvsimple}
\usepackage{enumitem}
\usepackage{fancybox}
\usepackage{fontenc}
\usepackage{graphicx}
\usepackage{listings}
\usepackage{longtable}
\usepackage{lscape}
\usepackage{makecell}
\usepackage{marvosym}
\usepackage{moreverb}
\usepackage{multicol}
\usepackage{multirow}
\usepackage{pifont}%
\usepackage{rotating}
\usepackage{setspace}
\usepackage[most]{tcolorbox}
\usepackage{threeparttable}
\usepackage{tikz}
\usepackage[normalem]{ulem}
\usepackage{url}
\usepackage{soul}
\usepackage{wasysym}
\usepackage{textcomp}
\usepackage{xcolor}
\usepackage{wrapfig}
\usepackage{pdflscape}
\usepackage{hyperref}

\usepackage{bigstrut}
\usepackage{balance}
\usepackage{xspace}

\newcolumntype{L}[1]{>{\raggedright\let\newline\\\arraybackslash\hspace{0pt}}m{#1}}
\newcolumntype{C}[1]{>{\centering\let\newline\\\arraybackslash\hspace{0pt}}m{#1}}
\newcolumntype{R}[1]{>{\raggedleft\let\newline\\\arraybackslash\hspace{0pt}}m{#1}}

\newboolean{showcomments}
\setboolean{showcomments}{true}
\ifthenelse{\boolean{showcomments}}
 { \newcommand{\mynote}[2]{
      \fbox{\bfseries\sffamily\scriptsize#1}
        {\small$\blacktriangleright$\textsf{\emph{#2}}$\blacktriangleleft$}}}
        { \newcommand{\mynote}[2]{}}

\newcommand{\datasetname}{MegaVul\xspace}

\usepackage{xcolor}
\usepackage[normalem]{ulem} 
\newcommand\hlm{\bgroup\markoverwith
  {\textcolor{yellow}{\rule[-.5ex]{2pt}{2.5ex}}}\ULon}

\copyrightyear{2024}
\acmYear{2024}
\setcopyright{acmlicensed}\acmConference[MSR '24]{21st International Conference on Mining Software Repositories}{April 15--16, 2024}{Lisbon, Portugal}
\acmBooktitle{21st International Conference on Mining Software Repositories (MSR '24), April 15--16, 2024, Lisbon, Portugal}
\acmDOI{10.1145/3643991.3644886}
\acmISBN{979-8-4007-0587-8/24/04}

\begin{document}

\title{\datasetname: A  C/C++ Vulnerability Dataset with Comprehensive Code Representations}

\author{Chao Ni, Liyu Shen, Xiaohu Yang, Yan Zhu}
\authornote{Both Chao Ni and Liyu Shen contribute equally. \\Yan Zhu is the corresponding author.}
\email{{chaoni,liyushen,yangxh,yan.zhu}@zju.edu.cn}
\affiliation{%
  \institution{Zhejiang University}
  \city{Hangzhou}
  \country{China}
}

\author{Shaohua Wang}
\email{davidshwang@ieee.org}
\affiliation{%
  \institution{Central University of Finance and Economics}
  \city{Beijing}
  \country{China}
}

\begin{abstract}
We constructed a newly large-scale and comprehensive C/C++ vulnerability dataset named \textbf{\datasetname} by crawling the Common Vulnerabilities and Exposures (CVE) database and CVE-related open-source projects. 
Specifically, we collected all crawlable descriptive information of the vulnerabilities from the CVE database and extracted all vulnerability-related code changes from 28 Git-based websites.
We adopt advanced tools to ensure the extracted code integrality and enrich the code with four different transformed representations.
Totally, \datasetname contains 17,380 vulnerabilities collected from 992 open-source repositories spanning 169 different vulnerability types disclosed from January 2006 to October 2023.
Thus, \datasetname can be used for a variety of software security-related tasks including detecting vulnerabilities and assessing vulnerability severity.
All information is stored in the JSON format for easy usage.
\datasetname is publicly available on GitHub and will be continuously updated.
It can be easily extended to other programming languages.
\end{abstract}



\maketitle

\input{sections/introduction}
\label{sec:introduction}

\input{sections/dataset}

\label{sec:dataset}

\input{sections/data_analysis}
\label{sec:data_analysis}

\input{sections/data_application}

\label{sec:data_application}

\input{sections/limitations}
\label{sec:limitations}

\input{sections/related_work}

\label{sec:related_work}

\input{sections/conclusion}
\label{sec:conclusion}

\vspace{-0.2cm}
\section*{Acknowledgements}{
This research is supported by the National Natural Science Foundation of China (No. 62202419), the Natural Science Foundation of Zhejiang Province (No. LY24F020008), and the Ningbo Natural Science Foundation (No. 2022J184).
}

\balance
\bibliographystyle{ACM-Reference-Format}
\bibliography{main}


\end{document}

%% file: sections/introduction.tex
\section{Introduction}

Data-driven software vulnerability analysis has been the core and critical activity in both industry and academia.
In particular, vulnerability detection attracts much attention from the research community~\cite{ni2023distinguish,wang2023deepvd,fu2022linevul,Wen2023vuldetect} since undetected vulnerabilities can be exploited by hackers and consequently lead to great losses to users. 
For example, a newly exploited vulnerability in a commonly used logging tool \textit{Apache Log4J} (\href{https://nvd.nist.gov/vuln/detail/CVE-2021-44228}{CVE-2021-44228}) makes millions of applications that adopt Log4J for logging execution information under attack.
Therefore, many studies proposed several datasets for better detecting software vulnerabilities.
However, existing datasets still have some limitations that may impact the verification of proposed models, including \ding{172} \textit{unreal vulnerability} (i.e., SARD~\cite{sard} is artificially synthesized), \ding{173} \textit{unreal data distribution} (i.e., balanced distribution in Devign~\cite{zhou2019devign}), \ding{174} \textit{limited diversity} (i.e., limited projects and vulnerability types in ReVeal~\cite{chakraborty2021deep}), \ding{175} \textit{limited newly disclosed vulnerabilities} (i.e., no updating to Big-Vul~\cite{fan2020ac} covering the period only from 2003 to 2019), and \ding{176} \textit{low-quality of dataset} (i.e., incomplete function, erroneously merged functions, missed commit message in Big-Vul~\cite{fan2020ac}).

Considering the aforementioned limitations, we constructed a new high-quality, data-rich, multi-dimensional C/C++ vulnerability dataset named \textbf{\datasetname} from the Common Vulnerabilities and Exposures (CVE) database~\cite{cve} and open-source projects. 
{First, we crawled the public CVE database to collect all of the available
descriptive information of a CVE (e.g., the CVE severity score, references linking to the affected products, etc). 
Second, through the CVE references directly or indirectly, we dug into the
relevant products hosted on Git-based websites (i.e., 28 in total) and then identified these vulnerability-related code commits and extracted relevant information (e.g., metadata, commit files).
We adopt advanced tools to ensure the extracted code integrality and to enrich the dimension of information.
In total, \datasetname contains 17,380 vulnerabilities collected from 992 open-source repositories spanning 169 different vulnerability types disclosed from January 2006 to October 2023.
Notice that the scripts for collecting \datasetname are publicly available and can be easily extended to other programming languages and implemented with more functionality.
\datasetname can be used for a variety of software security-related tasks, but not limited to, \ding{172} deep analysis on vulnerability characteristics and \ding{173} data-driven vulnerability detection and identification of vulnerability fixing patches.}
Eventually, the contributions of our paper are summarized as follows:

\textbf{[A. Continuously updating Dataset.]}
We collected and published a large-scale dataset with comprehensive basic information from both the CVE dataset and open-source project repositories as well as the enriched transformed representation of involved codes.

\textbf{[B. Open-soured Collection Process.]}{The data collection approach with supporting scripts is publicly available on GitHub~\cite{vul4c_github}.}

\begin{figure*}[!htbp]
\vspace{-0.3cm}
\centering
    \includegraphics[width=.85 \linewidth]{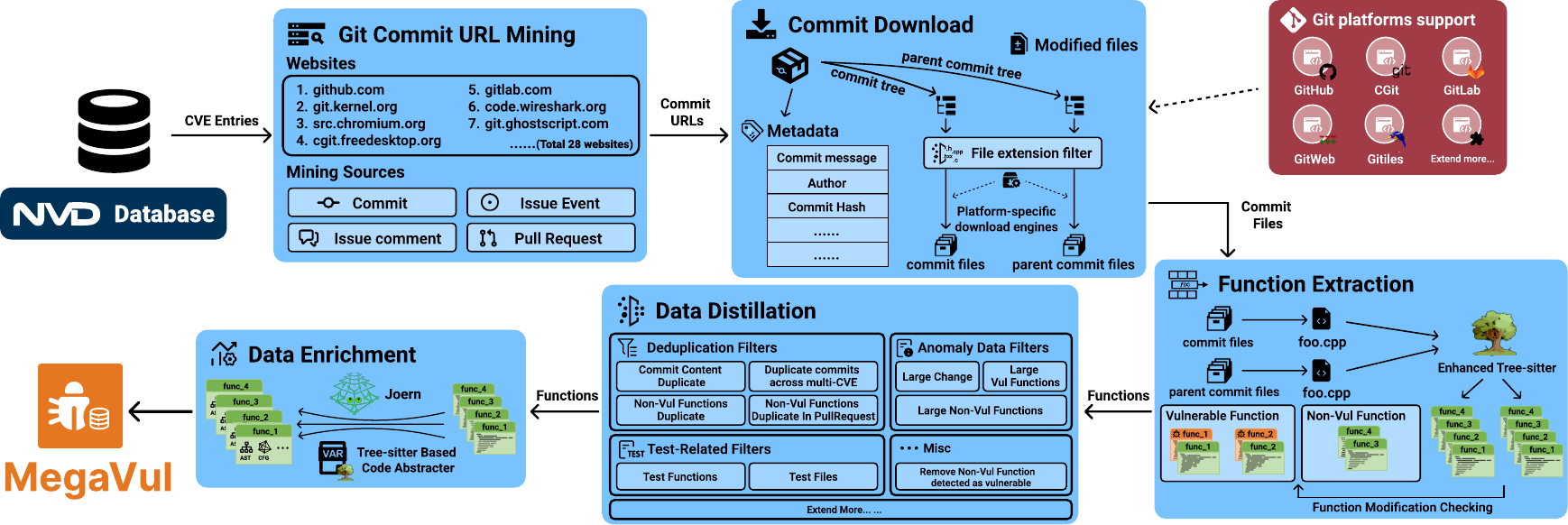}
    \caption{\datasetname data collection pipeline}
    \label{fig:vul4c_overview}
    \vspace{-0.3cm}
\end{figure*}

%% file: sections/dataset.tex
\section{Dataset Construction}



To address the aforementioned limitations in the previous dataset, we crawl data from more open-source repositories, using sophisticated filtering methods to improve the quality of the collected data. Furthermore, we employ advanced techniques to extract functions. 
Apart from collecting the raw functions, we also provide additional dimensions information of functions, including the abstracted versions and the information in the form of graphs obtained by \textit{Joern}~\cite{joern} such as the AST and the CFG of the functions.
This supplementary information avoids the re-implementation for those people who want to extract this information since it requires a relatively complex setup and consequently facilitates further research in this field. 
Table~\ref{tab:vul4c} presents the comparative statistical information between our dataset~\datasetname and Big-Vul.
Meanwhile, Figure~\ref{fig:vul4c_overview} depicts the workflow for data collection with five steps, and each step of the pipeline will be described in detail as follows.







\begin{table}[htbp]
\vspace{-0.3cm}
  \caption{Descriptive statistics for \datasetname and Big-Vul}
    \resizebox{\linewidth}{!}
    {
    \begin{tabular}{l|r|r}
    \toprule
 & \textbf{\datasetname} & \textbf{Big-Vul}\\
\midrule

Number of Repositories & \textbf{992} & 310\\
Number of CVE IDs &\textbf{ 8,254} & 3,539\\
Date range of crawled CVEs & \textbf{2006/01$\sim$2023/10} & 2013/01$\sim$2019/03 \\
Number of CWE IDs & \textbf{169} & 92\\
Number of Commits & \textbf{9,019} & 4,058\\
Number of Crawled Websites   &  \textbf{28} & 2 \\
Number of Vul/Non-Vul Function & \textbf{17,380/322,168} & 10,900/177,736  \\ 
Function Extract Method/(Quality)  & \textbf{Tree-sitter/(High)} &  Lizard/(Low) \\ 
Dimensions of Information & 
\textbf{6} & 2 \\
Code Integrality &\textbf{Full} & Partial \\
\bottomrule
\end{tabular}}
\label{tab:vul4c}
\vspace{-0.3cm}
\end{table}

\uline{\textbf{\ding{182} Git Commit URL Mining}}. 
The National Vulnerability Database (NVD)~\cite{nvd} is a public resource maintained by NIST to collect, organize, and distribute information on security vulnerabilities in software products, which is a good starting point for collecting a high-quality, comprehensive dataset for vulnerability detection. 
We crawl all available Common Vulnerabilities and Exposures (CVE)~\cite{cve} entries provided by the NVD into JSON format, which contains valuable descriptive information, including \textit{CVE summary}, \textit{CVSS scores}, and \textit{reference links to relevant products} or \textit{patches for fixing the vulnerabilities}.
However, the majority of references do not directly link to the URL of the software's commit but to the reference security bulletin or vulnerability advisory.
As a result, we conducted a manual analysis of the referenced commit links in CVE entries, identifying 28 Git-based code hosting platforms from 368 websites that had referenced the CVE entries more than 50 times.
Apart from identifying directed commit URLs, we also employed three other methods to mine potential commit URLs from the candidate websites:
\ding{172} \textit{Issue Event}: Locating the \textit{ClosedEvent} or \textit{CrossReferencedEvent} in the issue timeline that references commit URL;
\ding{173} \textit{Issue Comment}: Extracting the commit URL from specific comment formats within issues, such as the commit message comment auto-generated by the BOT;
\ding{174} \textit{Pull Request (PR)}: Identifying direct PR URL and mine PR URL from issues;

\uline{\textbf{\ding{183} Commit Download}}. 
In the previous step, the obtained commit URLs point to vulnerability-fixing commits.
Following previous work~\cite{chakraborty2021deep}, for each modified function in a vulnerability fix commit, the previous version of the function in the commit (i.e., the parent commit) is annotated as vulnerable, while the current version of the function is treated as non-vulnerable. 
Meanwhile, those functions that remain unchanged are also annotated as non-vulnerable.
Therefore, in this step, we download the metadata of commits as well as the files that have been modified compared to the parent commit. 
To minimize the size of downloading irrelevant data, we only gather C/C++-related source codes and headers based on their file extensions.
These commit URLs are sourced from various Git web hosting services, and to download commits from these platforms, we categorize these code hosting platforms into five main categories: GitHub, GitLab, GitWeb, CGit, and Gitiles. 
For each of these platforms, we implement the corresponding scripts to download commits.
For example, we utilize the official GitHub API to handle downloads from GitHub and employ a crawler for GitWeb to traverse the commit's metadata and extract the download path of the files.

\uline{\textbf{\ding{184} Function Extraction}}. 
Separating the functions in each file is a crucial step to ensure the quality and integrality of function extraction.
Existing datasets~\cite{alves2016software,sard,li2018vuldeepecker,zhou2019devign,fan2020ac} are collected through predefined rules or using regular expression tools like \textit{lizard}. However, these methods have several limitations in their functionalities due to their simplicity.
For example, regular expressions cannot fully represent the complex syntax rules of the C-like programming language. 
As a result, these methods may result in the extraction of incomplete functions or the incorrect separation of multiple functions into a single function.
Oppositely, we adopt a complex, syntax rule-based parser \textit{tree-sitter} to separate functions, 
and enhance the syntax rules to identify more macro modifiers frequently used by popular open-source repositories, such as the macro modifier \texttt{\_\_init} defined in Linux.
Furthermore, we temporarily remove the macros from the code before parsing, which allows the parser to effectively and accurately separate the functions without the interference of macros.


\begin{figure}[htbp]
\vspace{-0.2cm}
\centering
\includegraphics[width=\linewidth]{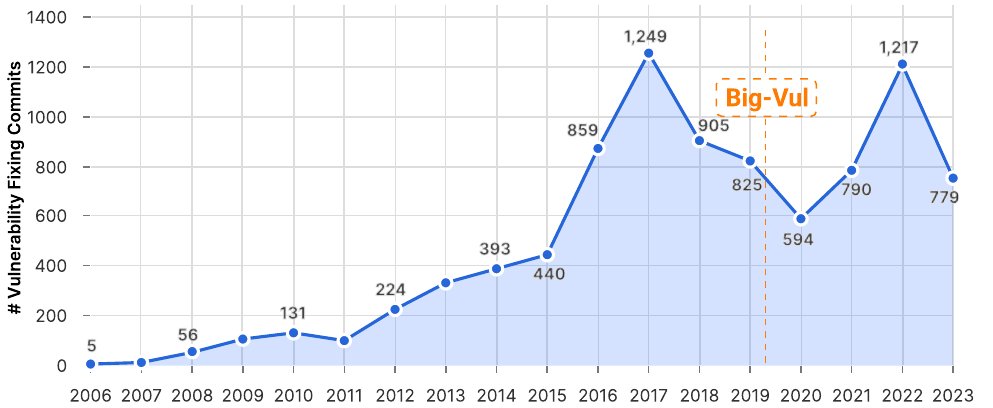}
\caption{Commit Count of Annual Vulnerability}
\label{fig:commits_over_year}
\vspace{-0.3cm}
\end{figure}

\uline{\textbf{\ding{185} Data Distillation}}. 
To address the noise in the raw dataset, multiple filters are applied to obtain the high-quality dataset, reducing data redundancy and leakage, and ensuring that the model trained with \datasetname demonstrates high performance, fairness, and reliability.
\ding{172} \textit{Deduplication Filters} aim to filter out duplicated functions to keep the dataset simple.
When a commit fixes multiple vulnerabilities and is referenced by multiple CVEs, we keep only one commit copy to ensure a single data source.
Deduplicating fixing commits with different \textit{hash} (i.e., an ID to indicate a commit) might be the same since they refer to the patches on different branches.
Deduplicate the same non-vulnerable functions across datasets.
Deduplicate identical non-vulnerable functions in multiple commits within a single pull request.
\ding{173} \textit{Anomaly Data Filters}:
To filter these templated codes generated by tools (i.e., \textit{yacc-generate}d or \textit{antlr-generated}), we adopt a 2-sigma criterion (i.e., 95\%) based on their length to filter both vulnerable or non-vulnerable functions.
We also remove large-scale changes including PR with many commits since a large number of modified files or diff lines usually arise from software major version bumps or code refactoring.
\ding{174} \textit{Test-Related Filters} aim to filter out files or functions related to unit testing based on their names.
\ding{175} \textit{Other Filters} aim to filter non-vulnerable functions that have been annotated as vulnerable after a specific commit to ensure label consistency and prevent the same function from being annotated as both vulnerable and non-vulnerable;

\uline{\textbf{\ding{186}  Data Enrichment}}.
In this step, we aim to enrich \datasetname information by providing more types of representation.
For example, graph representation is tedious work but extremely required by graph-based models~\cite{chakraborty2021deep,zhou2019devign,li2021vulnerability}.
``Abstraction'' (e.g., substituting variable names with symbolic names like ``VAR''~\cite{li2021sysevr}) is also needed by token-based sequence models to address the problem of vocabulary explosion.
Eventually, \datasetname provides information on the four dimensions of the function: 
\ding{172} \textit{Function Signature}: includes the function name, a comprehensive list of arguments with their respective names and types, and the return type of the function.
\ding{173} \textit{Abstracted Function}: provides nine types of granularity abstraction (e.g., FUNC, VAR, STRING, COMMENT, etc.).
\ding{174} \textit{Parsed Function}: means the functions with several types of representations (i.e., Abstract Syntax Trees (AST), Program Dependence Graphs (PDG)~\cite{cpg-ferrante1987program}, etc.) statically parsed by \textit{Joern}.
\ding{175} \textit{Code Changes}: include the details of function changes generated by \textit{diff} tool~\cite{git}.
For example, line-level modifications, additions, and deletions.

%% file: sections/data_analysis.tex
\section{Data Analysis}

Figure~\ref{fig:commits_over_year} clearly shows the trend in the number of C/C++-related vulnerability fixing commits per year since 2006. 
Overall, the number has increased dramatically since 2015, which indicates the aggressive vulnerability-fixing efforts by vendors and developers. 
The peak occurred in 2017 with 1,255 commits in total.
In particular, \textit{ImageMagick} project contributes the most, and the majority of vulnerability types are related to memory overflow vulnerabilities in different image formats, which led to a denial-of-service attack on remote servers and reported 357 CVEs in that year.
Despite a downward trend in subsequent years, the number of vulnerability fixing commits remains relatively high-level and arrive at another peak in 2022.
On the one hand, \textit{Vim} project reported 115 CVEs of buffer overflow vulnerabilities, which is much higher than the average number reported every year (i.e., about 10 CVEs).
On the other hand, the rise of deep learning has led to an increasing number of researchers who tend to use the TensorFlow framework developed with C++ backend operators to build deep learning models. 
Since 2021, TensorFlow has reported an average of 130 CVEs per year.
Notice that the most comprehensive dataset Big-Vul stopped collecting data in March 2019, but there are still about 4,000 commits related to vulnerability fixes between 2019 and 2023. 
Therefore, a continuously updating dataset is very beneficial for vulnerability detection tasks, which helps to richer the number and types of vulnerabilities and to learn deep learning-based detection models.

\begin{figure}[htbp]
\vspace{-0.3cm}
\centering
\includegraphics[width=\columnwidth]{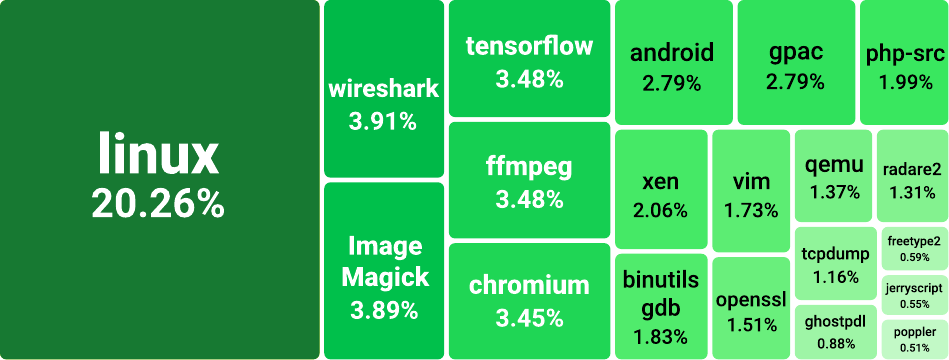}
\caption{Distribution of the Top 20 Repositories in \datasetname}
\label{fig:top20_repo_distribution}
\vspace{-0.3cm}
\end{figure}

Figure~\ref{fig:top20_repo_distribution} shows the Top-20 repositories with the highest number of CVEs in the \datasetname, which account for 59.5\% of the number of CVEs recorded by \datasetname, i.e., 4,914 CVE entries.
Especially, CVEs from \textit{Linux} account for 20.26\%, i.e., 1,672 CVEs.
On average, \datasetname collects 110 CVEs and their corresponding vulnerability fixing commits from \textit{Linux} every year.
In addition to \textit{Linux}, repositories such as \textit{Wireshark}, \textit{ImageMagick}, \textit{tensorflow}, \textit{ffmpeg}, and \textit{chromium} also show a relatively high percentage, with each repository being able to collect about 300 relevant CVEs. 
Different from Linux which has a stable and substantial trend in the annual amount of CVEs report, other repositories may intermittently report a significant number of CVEs in particular years.

\begin{figure}[htbp]
\vspace{-0.3cm}
\centering
\includegraphics[width=\linewidth]{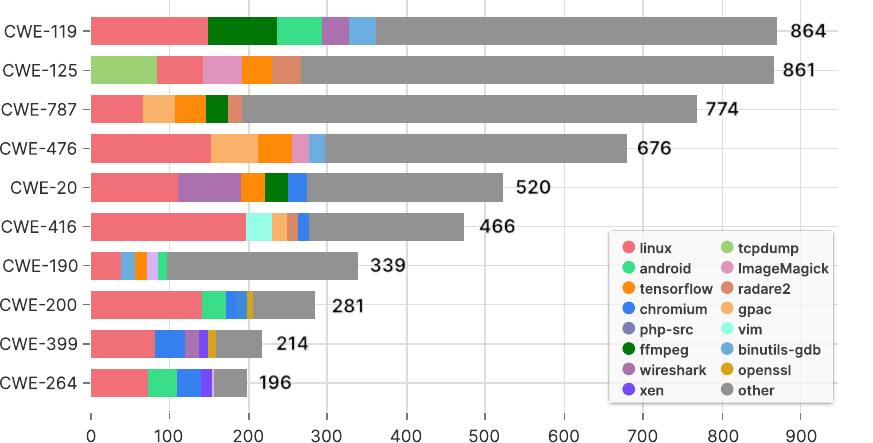}
\caption{The Top-10 CWEs and the Top-5 repositories with most frequent occurrences in each CWE}
\label{fig:top10_cwe}
\vspace{-0.3cm}
\end{figure}

Figure~\ref{fig:top10_cwe} illustrates the Top-10 CWE types in \datasetname as well as the Top-5 repositories with the highest frequency of each CWE type. 
In terms of the number of CWE types, CWEs related to memory management safety dominate (i.e., CWE-119, CWE-125, CWE-787, CWE-476, CWE-20, CWE-416, and CWE-190) account for 59.27\% of the total number of CWEs in \datasetname.
When analyzing each CWE type, we find that the \textit{Linux} has the highest number of vulnerabilities across multiple CWE types, except for CWE-125, which also highlights the diversity of vulnerabilities in the \textit{Linux}. 
Meanwhile, we find that Top-5 frequently occurring repositories have a wide distribution of vulnerability types which inspires us to pay attention to each category of vulnerabilities

\vspace{-0.2cm}

%% file: sections/data_application.tex
\vspace{-0.2cm}
\section{Data Application}

\datasetname can be used for a variety of software safety-related tasks because of its richness of information as well as its large-scale size.
We encourage researchers to use \datasetname as a benchmark to ensure a fair and uniform evaluation of SOTA performance and to avoid biased results caused by switching between different datasets. 

\textbf{The deep analysis on CVEs and code changes}.
\datasetname contains various types of textual data, such as CVE descriptions, commit messages, and code changes. 
By leveraging advanced NLP techniques (e.g., CodeBERT~\cite{ni2023distinguish}), we can analyze the correlations among these textual data,  understand the connotations of the code changes, identify the key information inside them, and dig deeper into the patterns of vulnerability.
This analysis not only helps to determine the location of the vulnerability and assess the possible impact of the vulnerability but also provides assistance in automated code review.

\textbf{Data-driven Vulnerability Detection}.
\datasetname covers a rich set of function information, including vulnerability functions, abstracted functions, function graphs, and other descriptive information such as the types of CWEs and CVE description. 
Researchers can utilize the rich information to train state-of-the-art deep learning models designed for vulnerability detection, including sequence-based and graph-based.
In addition, \datasetname is collected from several real-world repositories, covering functions with and without vulnerabilities with a realistic proportion.
Therefore, it supports training vulnerability detection models for better use in real-world production environments.

\textbf{Identification of vulnerability fixing patches}.
Due to the delay in disclosing vulnerability information after they have been patched, it creates a window of opportunity for attackers. 
Thus, many methods have been proposed for the identification of vulnerability-fixing commit and these methods can help downstream developers to fix vulnerabilities as early as possible.
\datasetname provides a large amount of commit metadata information and code change that can help train deep learning models to automatically identify vulnerability-fixing patches.

%% file: sections/limitations.tex
\section{LIMITATIONS}

During the collecting process, we manually selected data sources from websites with more than 50 CVE references, but some websites with less frequent references might still contain potential commits.
Besides, some projects use a different version control tool in early development, such as chromium using SVN a long time ago.
We do our best to find the corresponding SVN commits in repositories that have been migrated to Git, but some commits may still be missing.
During the de-duplication process, some functions with the same functionality may have differences in different branches of the codebase.
For example, a function may have compiler-specific macros in a new version of the codebase while the old version lacks it.
Although theoretically representing the same function, it is challenging to automatically determine whether their content is identical, resulting in duplication in the dataset.
In the future, we will adopt advanced techniques (i.e., NLP and refactoring detection techniques) to further improve data quality.

%% file: sections/related_work.tex
\vspace{-0.2cm}
\section{Related Work}

Several vulnerability databases for C/C++ have been previously proposed~\cite{alves2016software,sard,li2018vuldeepecker,zhou2019devign,fan2020ac}.
Alves et al.~\cite{alves2016software} collected vulnerability code from five widely used libraries.
They identified unique vulnerability identifiers (i.e., CVE IDs) and their corresponding commits from the security reporting websites of each library.
SARD~\cite{sard} artificially synthesizes vulnerable code by using well-known vulnerable patterns, which seems to be too simple to reflect the complexity of vulnerability patterns in the real world and is further extended by VulDeePecker~\cite{li2018vuldeepecker} with a focus on two specific CWE types (i.e., CWE-119 and CWE-399).
Both Devign~\cite{zhou2019devign} and Reveal~\cite{chakraborty2021deep} are also collected from several real-world open-source projects.
In addition, VulData7~\cite{vuldata7}, Big-Vul~\cite{fan2020ac}, and CVEfixes~\cite{bhandari2021cvefixes} collect vulnerabilities from NVD databases and find possible vulnerability fixing commits by referencing the links provided in reported CVEs.

%% file: sections/conclusion.tex
\vspace{-0.2cm}
\section{Conclusion}

We release a new dataset called \textbf{\datasetname}, which automatically extracts reference links from the NVD database for each CVE, mines potential vulnerability fixing commits, and downloads the commit metadata and files from the five popular Git-based web services. 
\datasetname utilizes an enhanced syntax rule-based \textit{Tree-sitter} to extract high-quality functions and apply several filters to enhance the dataset's quality.
Moreover, we have appended additional information to the functions, including abstract functions at nine different granularities, parsed function signatures, and various graph data, which significantly avoid duplication work on code pre-processing for downstream researchers.
Furthermore, we provide an interface to support more filters and Git-based platforms, and the implementation of \datasetname can be easily migrated to other programming languages (e.g., Java) with minor code modifications.

In conclusion, \datasetname has gathered high-quality functions from 9,019 commits, including 17,380 vulnerable and 322,168 non-vulnerable functions. 
All scripts and the dataset are publicly available on GitHub.
We will continue to update and expand the dataset to improve its quality and advance research in the field of security.